\documentclass[pteplogo]{ptephy_v1}

\usepackage{graphicx}
\usepackage{braket}
\usepackage{amsmath}
\usepackage{amssymb}
\usepackage{bm}

\preprintnumber{XXXX-XXXX}

\begin{document}

\title{Zero-energy states in conformal field theory with sine-square deformation}
\author[1,*]{Shota Tamura}
\author[1]{Hosho Katsura}
\affil{Department of Physics, Graduate school of Science, The University of Tokyo, 7-3-1 Hongo, Tokyo 113-0033, Japan\email{tamura@cams.phys.s.u-tokyo.ac.jp}}

\begin{abstract}
We study the properties of two-dimensional conformal field theories (CFTs) with sine-square deformation (SSD). We show that there are no eigenstates of finite norm for the Hamiltonian of a unitary CFT with SSD, except for the zero-energy vacuum state $\ket{0}$. 
We then introduce a regularized version of the SSD Hamiltonian which is related to the undeformed Hamiltonian via a unitary transformation corresponding to the {\it M$\ddot{o}$bius quantization}. The unitary equivalence of the two Hamiltonians allows us to obtain zero-energy states of the deformed Hamiltonian in a systematic way. The regularization also provides a way to compute the expectation values of observables in zero-energy states that are not necessarily normalizable. 
\end{abstract}

\maketitle

\section{Introduction}

Recently, 1+1 dimensional quantum many-body systems with sine-square deformation (SSD) have been investigated intensively. The SSD was first introduced by Gendiar {\it et al}. in \cite{SSD1} as a smooth boundary condition that efficiently suppresses boundary effects. 
A one-dimensional (1D) lattice model with the SSD has a local Hamiltonian density rescaled by the function $f(x)=\sin^2[\frac{\pi}{L}\left( x-\frac{1}{2} \right)]$, where $x$ is a position and $L$ is the length of the chain. This deformation leads to a system with open boundary conditions, i.e., the two end sites of the chain $x=1$ and $L$ are disconnected. Previous studies have revealed several remarkable properties of the SSD. 
In particular, for 1D quantum critical systems, the SSD leaves the ground state of the uniform Hamiltonian with periodic boundary conditions (PBC) almost unchanged. 
This correspondence between the systems with PBC and SSD was confirmed numerically in a variety of 1D critical systems, including free-fermion models, the extended Hubbard model, spin chains and ladders, and the Kondo lattice model \cite{SSD2, SSD3,  Shibata_Hotta, Hotta_Shibata, Hotta_Nishimoto_Shibata, Yonaga}. It was also proved analytically that the ground-state correspondence is exact for some models that are reducible to free fermions \cite{Example1, Example2, Example3, Katsura}.\par

One of the authors (HK) has shown in Ref. \cite{Katsura} that the underlying mechanism behind the ground-state correspondence can be understood in the framework of conformal field theories (CFTs) \cite{CFT_book1, CFT_book2}. 
In the language of CFT, the uniform and the SSD Hamiltonians are expressed as $\mathcal{H}_0=L_0+\bar{L}_0$ and 
\begin{equation}
\label{eq1}
\mathcal{H}_{\mathrm{SSD}}=L_{0}-\frac{L_1+L_{-1}}{2}+\bar{L}_0-\frac{\bar{L}_1+\bar{L}_{-1}}{2},
\end{equation}
respectively. Here, $L_n$ and $\bar{L}_n$, $n=0,\pm 1,\cdots$, are the generators of the Virasoro algebra. Conformally invariant vacuum states $\ket{0}$ and $\ket{\bar{0}}$ are defined to be annihilated by $L_m$ and $\bar{L}_m$ for $m\geq -1$, respectively. 
Therefore, the vacuum state $\ket{\mathrm{vac}}:=\ket{0}\bigotimes \ket{\bar{0}}$ is a common eigenstate of $\mathcal{H}_0$ and $\mathcal{H}_{\mathrm{SSD}}$ with eigenvalue $0$, i.e., 
$\mathcal{H}_0\ket{\mathrm{vac}}=\mathcal{H}_\mathrm{SSD}\ket{\mathrm{vac}}=0$. \par 

As shown by Ishibashi and Tada~\cite{Ishibashi}, the Hamiltonian (\ref{eq1}) has a continuous energy spectrum. 
For the convenience of the reader, we sketch their argument here. Take the coordinate transformation $u=\mathrm{e}^{2/(z-1)}$ where $z$ is the original complex coordinate. Then, with this complex coordinate $u$, we determine Virasoro generators via the radial quantization: $w=t+\mathrm{i}x=\log u$. Using an energy-momentum tensor $T(u)$, new Virasoro generators $\mathcal{L}_\kappa:=(2\pi \mathrm{i})^{-1}\oint_{|u|=\mathrm{const}} u^{\kappa+1}T(u)\mathrm{d}u$ can be defined uniquely for any real $\kappa$. Therefore, Virasoro generators
so obtained (called the {\it dipolar quantization}) has a continuous real index $\kappa$. One can confirm that $\mathcal{L}_0$ is expressed in terms of conventional Virasoro generators as $\mathcal{L}_0=L_0-(L_1+L_{-1})/2$. One can also show that the new generators $\mathcal{L}_n$'s satisfy the following commutation relation: 
$[\mathcal{L}_\kappa, \mathcal{L}_\kappa']=(\kappa-\kappa')\mathcal{L}_{\kappa+\kappa'}+(c/12)\kappa^3\delta(\kappa+\kappa')$. 
From the requirement that ${\displaystyle \lim_{t\to-\infty} T(z)\ket{0}=0}$, the vacuum state $\ket{0}$ must be annihilated by $\mathcal{L}_\kappa$ for all $\kappa>0$. Therefore, we have $\mathcal{L}_0(\mathcal{L}_{-\kappa}\ket{0}=\kappa(\mathcal{L}_{-\kappa}\ket{0})$ for $\kappa>0$, provided that ${\cal L}_0 \ket{0}=0$. The antiholomorphic part of $\mathcal{H}_\mathrm{SSD}$, $\bar{L}_0-(\bar{L}_1+\bar{L}_{-1})/2$, can be discussed similarly and hence $\mathcal{H}_\mathrm{SSD}$ has a continuous energy spectrum. \par

It has been revealed how the continuous energy spectrum in an SSD system is formed by considering the {\it M$\ddot{o}$bius quantization} \cite{Okunishi}. In this quantization, we take the coordinate transformation
\begin{equation}
\label{eq2}
w=-\frac{\sinh(\theta)-\cosh(\theta)z}{\cosh(\theta)-\sinh(\theta)z},
\end{equation}
where $\theta$ is an arbitrary real number, and calculate new Virasoro generators $\mathcal{L}_n(\theta)$ via the radial quantization with this coordinate. One can find that $\mathcal{L}_0(\theta)$ is represented as
\begin{equation}
\label{eq3}
\mathcal{L}_0 (\theta)=\cosh(2\theta)L_0-\sinh(2\theta)\frac{L_1+L_{-1}}{2},
\end{equation}
so that the M$\mathrm{\ddot{o}}$bius quantization with $\theta=0$ corresponds to the uniform system and that with $\theta \to +\infty$ corresponds to the SSD system except the normalization factor $\cosh(2\theta)$. The Hamiltonian $\mathcal{H}_{\mathrm{SSD}}(\theta)=[\mathcal{L}_0(\theta)+\bar{\mathcal{L}}_0(\theta)]/\cosh(2\theta)$ becomes $\mathcal{H}_\mathrm{SSD}$ as $\theta\to\infty$ and has a set of eigenvalues $n/\cosh(2\theta)$, $n=2,3,4,\cdots$. These eigenvalues can take all real non-negative values by setting $n/\cosh(2\theta)=\kappa$ and taking the limit $n \to \infty$ and $\theta \to \infty$, that is, the SSD limit. We note that a slightly different regularization of the SSD was discussed in~\cite{Wen_Ryu_Ludwig}.\par 

Though the relation between the uniform and the SSD systems was already established, most eigenstates found so far are not normalizable and it is hard to compute expectation values of physical quantities, such as the energy-momentum tensor $T(z)$. Therefore, we are left with two questions: (i) Is there a normalizable eigenstate in the CFT with the SSD? (ii) How should we calculate expectation values in the eigenstates whose norms diverge? To answer the first question, we first decompose the space of states into subspaces each of which is invariant under the action of $SL(2, \mathbb{R})$ generated by $\{L_0, L_{+1}, L_{-1} \}$. Then we show that the vacuum state $\ket{0}$ is the only normalizable eigenstate of the SSD Hamiltonian. Concerning the second question, we have an affirmative answer: the unitary transformation corresponding to the 
M$\mathrm{\ddot{o}}$bius quantization provides a regularization procedure that allows us to compute expectation values in the zero-energy states even though their norms diverge in the SSD limit. 

This paper is organized as follows. In Sec. 2, we introduce $SL(2,\mathbb{R})$ invariant subspaces of a Verma module. It is shown that there is no normalizable eigenstate of the SSD Hamiltonian, except for the vacuum state $\ket{0}$. In Sec. 3, we prove that each Virasoro operator $\mathcal{L}_n(\theta)$ in the M$\mathrm{\ddot{o}}$bius quantization can be written as a unitary transformation of $L_n$. In Sec. 4, we see that states obtained by considering the M$\mathrm{\ddot{o}}$bius quantization correspond to zero-energy states which were found in Ref. \cite{Tada} and their norms are finite for a finite $\theta$. In Sec. 5, we apply the technique to CFTs with other deformations to obtain their zero-energy states. 
We conclude with a summary in Sec. 6.

\section{Absence of eigenstates with finite norms}
\subsection{A Verma module}
In a 1+1 dimensional CFT, conformal symmetries of a system are described by Virasoro generators $L_n$ and $\bar{L}_n$, which form the Virasoro algebra
\begin{equation}
[L_n,L_m]=(n-m)L_{n+m}+\frac{c}{12}n(n^2-1)\delta_{n+m,0},
\end{equation}
where $c$ is a central charge\footnote{The antiholomophic part of the CFT can be discussed in the same manner as the holomophic part, and therefore we consider only the holomorphic part hereinafter.}. The $SL(2,\mathbb{C})$ invariant vacuum state is defined by the identity operator $I(z)$ as $\ket{0}:=I(z=0)$ and it obeys
\begin{equation}
L_n\ket{0}=0,\ \ n\ge -1.
\end{equation}
Throughout the paper, we only consider unitary CFTs. In a unitary CFT, a primary operator $\phi(z)$ has a non-negative conformal weight $h \ge 0$ and transforms under a conformal transformation $z\to w(z)$ as
\begin{equation}
\phi'(w)= \left( \frac{\mathrm{d}z}{\mathrm{d}w} \right)^h \phi(z).
\end{equation}
A primary state is then defined as $\ket{h}:=\phi(z=0)\ket{0}$ and it follows that
\begin{equation}
L_0\ket{h}=h\ket{h},\ \ L_n\ket{h}=0,\ \ n\ge 1.
\end{equation}
The hermitian conjugates of the states are defined as $\bra{0}:=I(z=\infty)$, ${\displaystyle \bra{h}:=\lim_{z\to\infty}z^{2h}\bra{0}\phi(z)}$, and $L_n^\dag=L_{-n}$. The factor $z^{2h}$ stems from the transformation
\begin{equation}
\phi'(w)=\phi(z) \left( \frac{\mathrm{d}w}{\mathrm{d}z} \right)^{-h} \propto \phi(z)z^{2h}
\end{equation}
under the coordinate transformation $w=1/z$.
The vacuum state is also one of the primary states because it is annihilated by $L_n$ with $n\ge 1$ and is the eigenstate of $L_0$ with eigenvalue 0.\par

\begin{figure}[h]
		\begin{center}
       \includegraphics{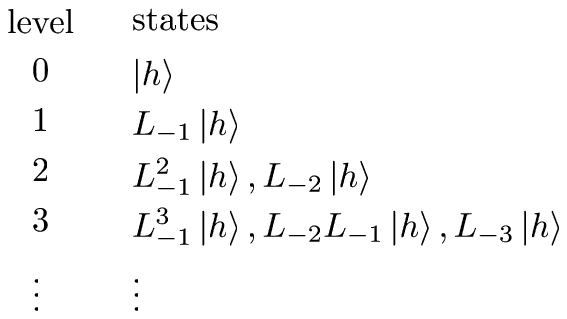}
      \caption{The Verma module}
		\label{fig3}
		\end{center}
\end{figure}

By acting with $L_{-n}$ ($n \ge 1$) on the primary states, one can construct a highest weight representation of the Virasoro algebra called a Verma module. The infinite set of states called ``descendants''
\begin{equation}
L_{-n_1}L_{-n_2}\cdots L_{-n_r}\ket{h}, \ \ n_1\ge n_2\ge \cdots \ge n_r>0
\end{equation}
build up a Verma module, and the integer $N:=\sum_{i=1}^r n_i$ is called the level of the state. A descendant with fixed $N$ is an eigenstate of $L_0$ with the eigenvalue $h+N$. Figure \ref{fig3} shows states which span the first few levels of the Verma module. \par

\subsection{$SL(2,\mathbb{R})$ invariant subspaces of Verma module}

A Verma module can be decomposed into $SL(2,\mathbb{R})$ invariant subspaces each of which is closed under the action of $L_0$ and $L_{\pm 1}$. Since the SSD Hamiltonian involves only $L_0$ and $L_{\pm 1}$, it is block diagonal with respect to these subspaces. In the following, we show how to construct such subspaces. 

Let $\ket{\psi^{(n)}_i} (i=1,2,\cdots,|\{ n \}|$) be orthonormal states at level $n$, where $|\{ n \}|$ denotes the number of states. One can recursively construct these states such that each state $\ket{\psi^{(n)}_i}$ is of the form $(L_{-1})^m \ket{\varphi}$, where $m=0,1,2,..., n$ and $\ket{\varphi}$ is a state that is annihilated by $L_1$. We prove this by mathematical induction. The case $n=0$ is clear, because there is only a primary state $\ket{h}$ annihilated by $L_1$. We now assume that the statement is true for level $n$, i.e., the states $\ket{\psi^{(n)}_i}$ form an orthonormal basis: 
\begin{equation}
\braket{\psi^{(n)}_k|\psi^{(n)}_i}=\delta_{k,i}, \quad\quad
k,i=1,2,\cdots,|\{ n \}|,
\end{equation}
and each is written in the form of $(L_{-1})^m \ket{\varphi}$ with $\ket{\varphi}$ annihilated by $L_1$. We can assume without loss of generality that each $\ket{\psi^{(n)}_i}$ is a simultaneous eigenstate of $L_0$ and the Casimir operator
\begin{equation}
{\bm J}^2 := L_{-1} L_1 +L_0 - (L_0)^2,
\end{equation}
which commutes with $L_0$ and $L_{\pm 1}$. We denote by $j_i (j_i -1)$ the eigenvalue of ${\bm J}^2$ in the state $\ket{\psi^{(n)}_i}$. Then it follows from the explicit calculation
\begin{align}
\braket{\psi^{(n)}_k | L_1 L_{-1} | \psi^{(n)}_i}
&= \braket{\psi^{(n)}_k | {\bm J}^2 + L_0 + (L_0)^2 | \psi^{(n)}_i} \nonumber \\
&= [ -j_i (j_i -1) +(h+n) (h+n+1) ] \braket{\psi^{(n)}_k | \psi^{(n)}_i},
\end{align}
that the states $L_{-1} \ket{\psi^{(n)}_i}$ ($i=1,\cdots,|\{ n \}|$) at level $n+1$ are orthogonal to each other. Therefore, the states defined by
\begin{equation}
\ket{\psi^{(n+1)}_i} = \frac{L_{-1} \ket{\psi^{(n)}_i}}{\sqrt{ -j_i (j_i-1) +(h+n) (h+n+1) }},
\quad i=1,2, ..., |\{ n \}|,
\end{equation}
form an orthonormal basis. Clearly, these states take the desired form. However, they do not exhaust all the states at level $n+1$ since $|\{ n \}| < |\{ n+1 \}|$ in a generic case. We denote by $\ket{\psi^{(n+1)}_i}$ ($|\{ n \}|+1\le i \le |\{ n+1 \}|$) the missing states orthogonal to $\ket{\psi^{(n+1)}_i}$ ($1\le i\le |\{ n \}|$) and assume that they are orthonormal. Note that this assumption eliminates the null states as their norms vanish. Then it is clear that the states $\ket{\psi^{(n+1)}_i}$ ($i=1,\cdots,|\{ n \}|$) form an orthonormal basis at level $n+1$. In this way, one can construct basis states, level by level, starting from the level $n=0$. 

It remains to prove that the states $\ket{\psi^{(n+1)}_i}$ ($|\{ n \}|+1\le i \le |\{ n+1 \}|$) are annihilated by $L_1$. The proof is as follows: Suppose for a contradiction that $L_1\ket{\psi^{(n+1)}_i}\neq 0$. Then it can be expanded as
\begin{equation}
L_1\ket{\psi^{(n+1)}_i}=\sum_{j=1}^{|\{n\}|}c_j \ket{\psi^{(n)}_j},
\end{equation}
where there is at least one $j$ for which the coefficient $c_j\neq 0$. Suppose $c_\ell \neq 0$. Then we find
\begin{equation}
\braket{\psi^{(n)}_\ell|L_1|\psi^{(n+1)}_i}=c_\ell\neq 0,
\end{equation}
which leads to a contradiction since we have assumed that $\ket{\psi^{(n+1)}_i}$ with $i= |\{ n \}|+1, ..., |\{ n+1 \}|$ is orthogonal to the states of the form $L_{-1}\,\ket{\psi^{(n)}_j}$ ($ 1 \le j \le |\{ n \}|$). This completes  the proof. 

\begin{figure}[h]
		\begin{center}
			\includegraphics[height=50truemm]{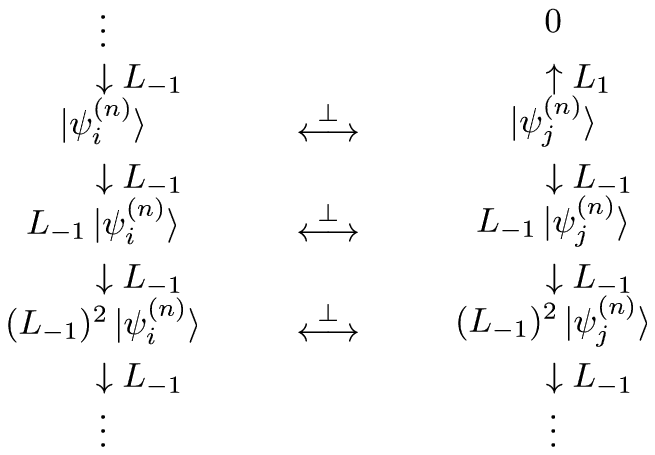}
      \caption{$SL(2,\mathbb{R})$ invariant subspaces of the Verma module}
		\label{fig4}
		\end{center}
\end{figure}

The above procedure yields the decomposition of the Verma module into $SL(2, \mathbb{R})$ invariant subspaces. In each subspace, the highest-weight state is the state that is annihilated by $L_1$. One can construct higher level states by acting with $L_{-1}$ on the highest-weight state repeatedly, as schematically shown in Fig. \ref{fig4}. We note that the vacuum $\ket{0}$ is special in that it is annihilated by both $L_1$ and $L_{-1}$, and hence is a trivial (one-dimensional) representation of $SL(2, \mathbb{R})$.\par

\subsection{Hamiltonian in each subspace}

Let us now discuss how the SSD Hamiltonian $\mathcal{H}_\mathrm{SSD}=L_0-(L_1+L_{-1})/2$ can be expressed in each subspace. Let $\ket{\varphi_0}$ be the highest-weight state of one of the subspaces, which is normalized as $\langle \varphi_0 | \varphi_0 \rangle=1$. Suppose the level of this state is $n$, i.e. $L_0\ket{\varphi_0}=(h+n)\ket{\varphi_0}$. The set of normalized states can be built from $\ket{\varphi_0}$ as
\begin{equation}
\begin{split}
\ket{\varphi_m }&:=\frac{1}{\sqrt{f_m}}L_{-1}\ket{\varphi_{m-1}},  
\quad m=1,2,\cdots,
\end{split}
\end{equation}
where
\begin{equation}
f_m=m(2h+2n+m-1),
\end{equation}
gives the normalization of the state $(L_{-1})^m\ket{\varphi_0}$. We note that states at different levels are orthogonal as they differ in their $L_0$ eigenvalues. 

The matrix elements of the SSD Hamiltonian, $\braket{\varphi_m|\mathcal{H}_\mathrm{SSD}| \varphi_{m'}}$, are nonvanishing only when $m'=m\ \mathrm{or}\ m\pm 1$, and one finds
\begin{align}
\braket{\varphi_{m} |\mathcal{H}_\mathrm{SSD}| \varphi_{m} }&=\braket{m|L_0|m}=h+n+m, \\
\label{19}
\braket{ \varphi_{m} |\mathcal{H}_\mathrm{SSD}| \varphi_{m+1}}&=\braket{\varphi_{m+1} |\mathcal{H}_\mathrm{SSD}|\varphi_{m} }=-\frac{1}{2}\braket{\varphi_{m}|L_1|\varphi_{m+1}}=-\frac{1}{2}\sqrt{f_{m+1}}.
\end{align}
Therefore, the Hamiltonian in the subspace takes the form
\begin{equation}
\mathcal{H}^{({\rm sub})}_\mathrm{SSD}=
\begin{pmatrix} 
\label{hssd}
h+n & -\sqrt{f_1}/2 &  & & \\
-\sqrt{f_1}/2 & h+n+1 & -\sqrt{f_2}/2 & & \\
 & -\sqrt{f_2}/2 & h+n+2 & -\sqrt{f_3}/2 & \\
 & & -\sqrt{f_3}/2 & h+n+3 & \ddots \\
 & & & \ddots & \ddots \\
\end{pmatrix},
\end{equation}
where matrix elements which are zero are left empty. Any eigenstate of $\mathcal{H}^{({\rm sub})}_{\rm SSD}$ with eigenvalue $\lambda$ can be expressed in terms of $\ket{\varphi_m}$ as $\ket{\lambda} = \sum_m c_m\ket{m}$, where $c_m$ are coefficients. 
The Schr{\"o}dinger equation, $\mathcal{H}^{({\rm sub})}_{\rm SSD} \ket{\lambda} = \lambda \ket{\lambda}$, for the components reads
\begin{eqnarray}
\label{eigeneq}
-\frac{1}{2}\sqrt{f_m} \, c_{m-1}+(h+n+m)c_m-\frac{1}{2}\sqrt{f_{m+1}} \, c_{m+1}=\lambda c_m, \ m=0,1,2,\cdots,
\end{eqnarray}
where we have set $c_{-1}=0$.

\subsection{Uniqueness of the eigenstate with a finite norm}
So far we have discussed the representation of the SSD Hamiltonian, and showed how one can determine its eigenstates. Using the result obtained, we will see that there are no eigenstates of finite norm, except the vacuum $\ket{0}$ which is a zero-energy state of $\mathcal{H}_{\rm SSD}$.

First, let us prove the absence of an eigenstate having a non-zero eigenvalue and finite norm. Suppose that the state $\ket{\lambda}$ satisfies $\braket{\lambda|\lambda}<\infty$ and $\mathcal{H}_\mathrm{SSD}\ket{\lambda}=\lambda \ket{\lambda}$. Then, we can normalize the state $\ket{\lambda}$ as $\braket{\lambda|\lambda}=1$. The eigenvalue $\lambda$ is real since the Hamiltonian $\mathcal{H}_\mathrm{SSD}$ is hermitian.  
Put $L_{-}:=(L_1-L_{-1})/2$. We have
\begin{equation}
[\mathcal{H}_\mathrm{SSD},L_{-}]=\biggl[ L_0-\frac{L_1+L_{-1}}{2},\frac{L_1-L_{-1}}{2} \biggr] = \mathcal{H}_\mathrm{SSD}.
\end{equation}
Then the expectation value of the commutator in $\ket{\lambda}$ gives 
\begin{equation}
\label{comm}
\lambda \braket{\lambda|L_{-}|\lambda}-\braket{\lambda|L_{-}|\lambda}\lambda=\lambda\braket{\lambda|\lambda},
\end{equation}
where the LHS is zero\footnote{If $\braket{\lambda|L_{-}|\lambda}$ diverges, Eq. (\ref{comm}) has no meaning. The discussion in the previous section shows that this problem does not occur in our case: Expand the eigenstate $\ket{\lambda}=\sum_m c_m\ket{m}$. From Eq. (\ref{eigeneq}), we find that the coefficients can be written as $c_m=d_m\, c_0, d_m\in \mathbb{R}$. Therefore, using Eq. (\ref{19}), we have
\begin{align}
\braket{\lambda|L_-|\lambda}&=\sum_{m=0}^\infty \left( \bar{c}_mc_{m+1} \frac{1}{2}\sqrt{f_{m+1}} -\bar{c}_mc_{m-1}\frac{1}{2}\sqrt{f_m} \right) 
=0. \nonumber
\end{align}
}. Therefore, the eigenvalue $\lambda$ must be zero from the assumption.\par

As we have seen, the vacuum $\ket{0}$, a trivial representation of $SL(2,\mathbb{R})$, is a zero-energy eigenstate of $\mathcal{H}_\mathrm{SSD}$. 
Let us prove that $\ket{0}$ is the unique zero-energy state with a finite norm. Suppose for the sake of contradiction that we can find another state $\ket{\psi_0}$ in some subspace that is a zero-energy state. Then $\ket{\psi_0}$ can be written as
\begin{equation}
\ket{\psi_0}=\sum_{m=0}^\infty c_m \ket{\varphi_m},
\end{equation}
and we can determine $c_m$ from $c_0$ via Eq. (\ref{eigeneq}) with $\lambda=0$. One can solve the recursion easily and get
\begin{equation}
c_m=\frac{1}{m!}\sqrt{f_1f_2\cdots f_m}, 
\quad m=1,2,\cdots.
\end{equation}
This can be proved by mathematical induction. 

We can prove that $\ket{\psi_0}$ is unnormalizable by direct calculation. Put $x:=2h+2n\ge 0$. Then we have
\begin{align}
\braket{\psi_0|\psi_0}&=\biggl[ 1+\sum_{m=1}^\infty \frac{(x)_m}{m!} \biggr] |c_0|^2 \nonumber \\
&=\biggl[ 1+\sum_{m=1}^\infty \frac{x}{m}\frac{(x+1)\cdot (x+2)\cdot \cdots \cdot (x+m-1)}{1\cdot 2\cdot \cdots (m-1)} \biggr] |c_0|^2 \nonumber \\
&\ge \biggl[ 1+\sum_{m=1}^\infty \frac{x}{m} \biggr] |c_0|^2, 
\end{align}
where the Pochhammer symbol is defined by $(x)_0:=1$ and $(x)_n:=x(x+1)\cdot\cdots\cdot(x+n-1)$ for $x\in \mathbb{C}$ and $n\in\mathbb{N}$. Since the series $\sum_{m=1}^\infty 1/m$ diverges, $\braket{\psi_0|\psi_0}$ diverges if $x=2h+2n>0$. Therefore, no normalizable zero-energy state exists except for $h=n=0$, in which case $\ket{\psi_0}$ is by definition the same as the vacuum $\ket{0}$.

\section{Equivalence of a unitary transformation and the M$\mathrm{\ddot{o}}$bius quantization}
\subsection{M$\ddot{o}$bius transformation written by Witt generators}

We have found that there is no normalizable eigenstate of the SSD Hamiltonian except for $\ket{0}$. A natural question is: How can one calculate expectation values of physical observables if one obtains an eigenstate of ${\cal H}_{\rm SSD}$ with a divergent norm? 
In fact, unnormalizable eigenstates were found so far \cite{Tada}. One answer to this problem is to take some limiting process. To this end, we introduce a one-parameter family of unitary transformations corresponding to the M$\ddot{\rm o}$bius quantization.\par    
It was shown by Matone that if operators $X$,$Y$, and $Z$ satisfy $[X,Y]=uX+vY+cI$ and $[Y,Z]=wY+zZ+dI$, there exist some c-numbers $a$ and $b$ such that
\begin{equation}
\exp(X)\exp(Y)\exp(Z)=\exp(aX+bY+c[X,Z]+dI),
\end{equation}
where $I$ is an identity operator \cite{Matone}. This formula was then applied to the Virasoro algebra 
\begin{equation}
[L_n,L_m]=(n-m)L_{n+m}+\frac{c}{12}n(n^2-1)\delta_{n+m,0}
\end{equation}
with $c$ the central charge, and led to
\begin{equation}
\begin{split}
\label{eq4}
&\exp(\lambda_{-k}L_{-k})\exp(\lambda_0L_0)\exp(\lambda_kL_k)=  \\ 
&\ \ \ \ \ \ \ \exp \biggl\{ \frac{\lambda_{+}-\lambda_{-}}{\mathrm{e}^{k\lambda_{-}}-\mathrm{e}^{k\lambda_{+}}} \biggl[ -k\lambda_{-k}L_{-k}+(2-\mathrm{e}^{k\lambda_{+}}-\mathrm{e}^{k\lambda_{-}})L_0-k\lambda_kL_k+c_kI \biggr] \biggr\}.
\end{split}
\end{equation}
Here $k$ is an arbitrary integer and $\lambda_{\pm k}$ and $\lambda_0$ are arbitrary real numbers. The constants $c_k$ and $\lambda_\pm$ are determined by
\begin{equation}
\begin{split}
c_k&=\frac{\lambda_{-k}\lambda_k}{\lambda_{+}-\lambda_{-}}\left( \frac{\lambda_{+}}{1-\mathrm{e}^k\lambda_{-}}-\frac{\lambda_{-}}{1-\mathrm{e}^{k\lambda_{-}}} \right) \frac{c}{12}(k^4-k^2), \\
\mathrm{e}^{k\lambda_{\pm}}&=\frac{1+\mathrm{e}^{k\lambda_0}-k^2\lambda_{-k}\lambda_k\pm \sqrt{(1+\mathrm{e}^{k\lambda_0}-k^2\lambda_{-k}\lambda_k)^2-4\mathrm{e}^{k\lambda_0}}}{2}.
\end{split}
\end{equation}
Applying this formula with $k=1$, $\lambda_{1}=\tanh(\theta)$, $\lambda_{-1}=-\tanh(\theta)$, and $\lambda_0=-\log[\cosh^2(\theta)]$, we have
\begin{eqnarray}
\label{eq5}
&\exp[-\tanh(\theta)L_{-1}]\exp[-L_0\log(\cosh^2(\theta))]\exp[\tanh(\theta)L_1]=\mathrm{e}^{\theta(L_1-L_{-1})}.
\end{eqnarray} 

By using the formula, the transformation (\ref{eq2}) can be written in a simple form. Because we can rewrite Eq. (\ref{eq2}) as the three consecutive $SL(2,\mathbb{R})$ transformations
\begin{equation}
\label{wmobius}
w=\frac{1}{\cosh^2(\theta)}\frac{z}{1-tz}-t
\end{equation}
where $t:=\tanh(\theta)$, this can be expressed in terms of the Witt algebra $\ell_n=-z^{n+1}\partial_z$\ (see Appendix. A):
\begin{equation}
\label{wuni}
w=\mathrm{e}^{-t\ell_1}\mathrm{e}^{\ell_0\log[\cosh^2(\theta)]}\mathrm{e}^{t\ell_{-1}}z.
\end{equation}
Since 
the two subalgebras $\{L_{-1},L_0,L_{1}\}$ and $\{\ell_{-1},\ell_0,\ell_1\}$ satisfy the same commutation relation, Eq. (\ref{wuni}) simplifies to
\begin{equation}
\label{eq6}
w=\mathrm{e}^{\theta(\ell_1-\ell_{-1})}z.
\end{equation}
From this, we expect that the Virasoro generators  $\mathcal{L}_n(\theta)$ in the M$\mathrm{\ddot{o}}$bius quantization can be obtained from $L_n$ via the unitary transformation generated by $L_1-L_{-1}$, i.e., 
$\mathcal{L}_n(\theta)=\mathrm{e}^{-\theta(L_1-L_{-1})}L_n\mathrm{e}^{\theta(L_1-L_{-1})}$. We prove this conjecture in the next section. 
\par

\subsection{A unitary transformation of Virasoro operators}
In Ref. \cite{Okunishi}, Virasoro operators with new coordinate $w$ is calculated as
\begin{equation}
\label{eq7}
\mathcal{L}_n(\theta)=(-1)^{n+1}\frac{\sinh(2\theta)}{2t^n}\sum_m L_m \oint_\tau \frac{\mathrm{d}z}{2\pi\mathrm{i}}\frac{(z-t)^{n+1}}{(z-1/t)^{n-1}}z^{-m-2},
\end{equation}
and the relation $\mathcal{L}_{-n}(\theta)=\mathcal{L}_n^\dag(\theta)$ is proved. Here, the integration path is taken so that the parameter $\tau$ defined by
\begin{equation}
\coth(\theta)\mathrm{e}^\tau=\left|\frac{\sinh(\theta)-\cosh(\theta)z}{\cosh(\theta)-\sinh(\theta)z}\right|,
\end{equation}
is kept constant. Suppose that the integration contour encloses the origin but does not include $1/t$. Then one can verify that Eq. (\ref{eq7}) can be rewritten as
\begin{equation}
\label{eq8}
\mathcal{L}_n(\theta)=(-1)^{n+1}\frac{\sinh(2\theta)}{2}\sum_{m=-1}^{\infty} \sum_{j=0}^{\min(m+1,n+1)} (-1)^j \frac{(n-j+2)_j(n-1)_{m-j+1}}{j!(m-j+1)!}t^{n-2j+m+1}L_m
\end{equation}
for $n\geq-1$ (for a derivation, see Appendix. B). For example, we have
\begin{equation}
\label{e.g.}
\mathcal{L}_2(\theta)=-\frac{\sinh(2\theta)}{2}\biggl[ t^2L_{-1}+(t^3-3t)L_0+(t^4-3t^2+3)L_1+\sum_{m=2}^\infty t^{m-3}(t^2-1)^3L_m \biggr].
\end{equation}
The operators $\mathcal{L}_n(\theta)$ for $n\leq -2$ can be obtained from the relation $\mathcal{L}_{-n}(\theta)=\mathcal{L}^\dag_n(\theta)$.\par 

Equation (\ref{eq6}), on the other hand, implies that the transformation of Virasoro operators can be written as the unitary transformation
\begin{equation}
\label{eq9}
\mathcal{L}_n(\theta)=\mathrm{e}^{-\theta(L_1-L_{-1})}L_n\mathrm{e}^{\theta(L_1-L_{-1})}.
\end{equation}
In the following, we will show the equivalence of Eq. (\ref{eq6}) and Eq. (\ref{eq9})\par

We first note that the RHS of Eq. (\ref{eq9}) contains only linear terms in $L_n$. This fact can be shown by using Baker-Campbell-Hausdorff's formula
\begin{equation}
\mathrm{e}^{A}B\mathrm{e}^{-A}=B+[A,B]+\frac{1}{2!}[A,[A,B]]+\frac{1}{3!}[A,[A,[A,B]]]+\cdots.
\end{equation}
Therefore, Eq. (\ref{eq9}) can be cast into the form
\begin{equation}
\label{eq10}
\mathcal{L}_n(\theta)=\frac{\sinh(2\theta)}{2}\sum_m \sum_{k=-\infty}^\infty y_n^{(m,k)}t^kL_m.
\end{equation}
The initial condition 
\begin{equation}
\label{condition}
\mathcal{L}_n(\theta=0)=L_n
\end{equation}
and $\sinh(2\theta)/2=t\cosh^2(\theta)$ give
\begin{equation}
\begin{cases}
\label{eq12}
y_n^{(m,k)}=0, & k\leq-2 \\
y_n^{(m,-1)}=\delta_{n,m} & 
\end{cases}.
\end{equation} \par
Secondly, by substituting Eq. (\ref{eq10}) to the derivative of Eq. (\ref{eq9}),
\begin{equation}
\frac{\mathrm{d}}{\mathrm{d}\theta}\mathcal{L}_n(\theta)=(n-1)\mathcal{L}_{n+1}(\theta)-(n+1)\mathcal{L}_{n-1}(\theta),
\end{equation}
we have
\begin{equation}
\label{eq13}
\begin{split}
\cosh(2\theta)\sum_{mk}&y_n^{(m,k)}t^kL_m+\frac{\sinh(2\theta)}{2\cosh^2(\theta)}\sum_{mk}ky_n^{(m,k)}t^{k-1}L_m \\
&=\frac{\sinh(2\theta)}{2}\sum_{mk}(n-1)y_{n+1}^{(m,k)}t^kL_m-\frac{\sinh(2\theta)}{2}\sum_{mk}(n+1)y_{n-1}^{(m,k)}t^kL_m.
\end{split}
\end{equation}
Using $2/\tanh(2\theta)=t+1/t$, the comparison of the coefficient of $t^kL_m$ in Eq. (\ref{eq13}) reads
\begin{equation}
\label{eq14}
(k+1)y_n^{(m,k)}-(k-3)y_n^{(m,k-2)}=(n-1)y_{n+1}^{(m,k-1)}-(n+1)y_{n-1}^{(m,k-1)}.
\end{equation}
We equivalently deformed Eq. (\ref{eq9}) to Eqs. (\ref{eq10}), (\ref{eq12}) and (\ref{eq14}) above. Finally, we show that Eq. (\ref{eq8}) is equivalent to Eqs. (\ref{eq10}), (\ref{eq12}) and (\ref{eq14}). \par

\begin{figure}[h]
	\begin{minipage}{0.5\hsize}
		\begin{center}
			\includegraphics[height=40truemm]{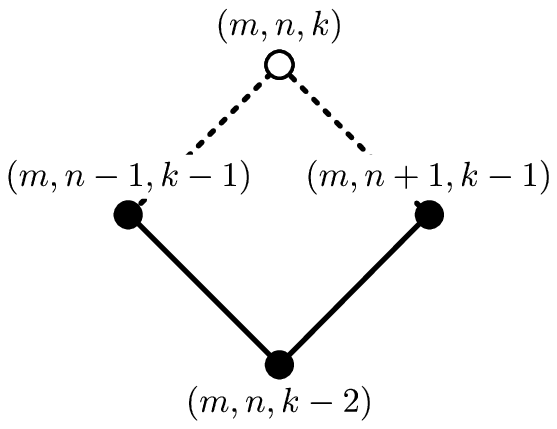}
      \caption{The meaning of Eq. (\ref{eq14}).}
		\label{fig1}
		\end{center}
	\end{minipage}
	\begin{minipage}{0.5\hsize}
		\begin{center}
			\includegraphics[height=40truemm]{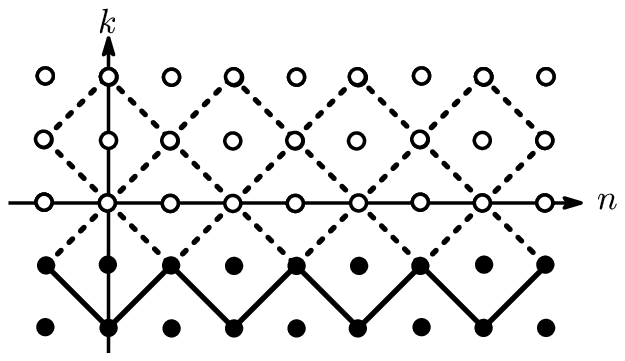}
		\caption{The value of $y_n^{(m,k)}$ is determined from the bottom line ($k=-1$) to the top in a one-by-one manner. 
}
		\label{fig2}
		\end{center}
	\end{minipage}
\end{figure}

Equations (\ref{eq12}) and (\ref{eq14}) determine each of the terms $y_n^{(m,k)}$ uniquely: One can think of $y_n^{(m,k)}$ as a function of $(m,n,k)$ in the cubic lattice $\mathbb{Z}^3$. Since the recursion relation Eq. (\ref{eq14}) is closed within a plane in which $m$ is constant, we can treat the problem for each $m$ separately. 
Equation (\ref{eq14}) also implies that the values at three points $(m, n-1, k-1)$, $(m, n+1, k-1)$, and $(m, n, k-2)$ give that at $(n, m, k)$ as shown in Fig. \ref{fig1}. Then, Fig. \ref{fig2} shows that the values of $y_n^{(m,k)}$'s for all $n$ and $k\leq -1$, fixed by the initial conditions Eq. (\ref{eq12}), determine the other terms uniquely. 

One thing left to do is to check if the coefficients of Eq. (\ref{eq8}) satisfy Eq. (\ref{eq14}). Putting $k=n+m-2j+1$, the coefficient of each term in the sum corresponds to $y_n^{(m,k)}$. If $(n+m)-k$ is an even integer, $y_n^{(m,k)}=0$ follows from $k=n+m-2j+1$ and Eq. (\ref{eq14}) is obviously satisfied. If $(n+m)-k$ is an odd integer, Eq. (\ref{eq8}) gives
\begin{equation}
\label{eq15}
y_n^{(m,k)}=(-1)^{\frac{n+m-k+1}{2}+n+1}\frac{(n-\frac{n+m-k+1}{2}+2)_{\frac{n+m-k+1}{2}}(n-1)_{m-\frac{n+m-k+1}{2}+1}}{(\frac{n+m-k+1}{2})!(m-\frac{n+m-k+1}{2}+1)!}.
\end{equation}
A tedious but straightforward calculation shows that the above $y_n^{(m,k)}$ satisfies Eq. (\ref{eq14}). 
Equation (\ref{eq15}) also satisfies Eq. (\ref{eq12}) because Eq. (\ref{eq8}) clearly fulfills the condition (\ref{condition}). Therefore, considering that Eq. (\ref{eq9}) clearly satisfies ${\cal L}_{-n} (\theta) = {\cal L}^\dagger_n (\theta)$, the equivalence between Eq. (\ref{eq8}) and (\ref{eq9}) has been shown for all $n$.

\section{Zero-energy states in a concrete form}
In the previous section, we saw that the M$\mathrm{\ddot{o}}$bius quantization can be thought of as a unitary transformation that relates $L_n$ and $\mathcal{L}_n(\theta)$ via Eq. (\ref{eq9}). Using this relation, we can construct a primary state in the CFT with the M$\mathrm{\ddot{o}}$bius quantization or zero-energy states of the CFT with the SSD in a natural form.

\subsection{Vacuum state in CFT with M$\ddot{o}$bius quantization}
As we have seen in the previous section, in the CFT constructed with the radial quantization, the vacuum state is defined as
\begin{equation}
\label{vacuum}
\ket{0}=I(z=0)\ket{0},\ \ \ L_n\ket{0}=0 \ \ \mathrm{for}\ n\geq -1,
\end{equation}
where $I(z=0)$ is the identity operator. \par
Then, the vacuum state in the CFT with the M$\mathrm{\ddot{o}}$bius quantization can be naturally defined as
\begin{align}
&\ket{0}_\theta=\mathrm{e}^{-\theta(L_1-L_{-1})}I(z=0)\mathrm{e}^{\theta(L_1-L_{-1})}\ket{0},\\
&\mathcal{L}_n(\theta)\ket{0}_\theta=0,\ \ n\geq -1.
\end{align}
This definition is the same as the definition (\ref{vacuum}) of $\ket{0}$ because of $L_1\ket{0}=L_{-1}\ket{0}=0$. Therefore, we find that the CFT with the radial quantization and the one with the M$\mathrm{\ddot{o}}$bius quantization have the same vacuum state: $\ket{0}_\theta=\ket{0}$.

\subsection{Primary state in CFT with M$\ddot{o}$bius quantization}
In the CFT with the M$\mathrm{\ddot{o}}$bius quantization, the primary state $\ket{h}_\theta$ can also be defined naturally as
\begin{align}
\label{def.0}
\ket{h}_\theta&=\mathrm{e}^{-\theta(L_1-L_{-1})}\phi(z=0)\mathrm{e}^{\theta(L_1-L_{-1})}\ket{h}, 
\end{align}
where $\ket{h}=\ket{h}_{\theta=0}$ is a primary state of the uniform Hamiltonian $L_0$. From the definition, we have
\begin{equation}
\mathcal{L}_0(\theta)\ket{h}_\theta=\mathrm{e}^{-\theta(L_1-L_{-1})}L_0\mathrm{e}^{\theta(L_1-L_{-1})}\mathrm{e}^{-\theta(L_1-L_{-1})}\ket{h}=h\mathrm{e}^{-\theta(L_1-L_{-1})}\ket{h}=h\ket{h}_\theta
\end{equation}
and ${}_\theta \braket{h|h}_\theta=1$ for all $\theta$. \par
This definition is consistent with that of Ref. \cite{Okunishi},
\begin{equation}
\ket{h}'_\theta=\mathrm{e}^{\tanh(\theta)L_{-1}}\ket{h}.
\end{equation}
Using Eq. (\ref{eq5}), we find
\begin{align}
\label{primary}
\ket{h}_\theta&=\mathrm{e}^{-\theta(L_1-L_{-1})}\ket{h} \nonumber \\
&=\mathrm{e}^{\tanh(\theta)L_{-1}}\mathrm{e}^{-L_0\log[\cosh^2(\theta)]}\mathrm{e}^{-\tanh(\theta)L_1}\ket{h} \nonumber \\
&=\frac{1}{\cosh^{2h}(\theta)}\mathrm{e}^{\tanh(\theta)L_{-1}}\ket{h} \nonumber \\
&=\frac{1}{\cosh^{2h}(\theta)}\ket{h}'_\theta.
\end{align}
Therefore, 
$\ket{h}_\theta$ is identical to the state $\ket{h}'_\theta$ up to the normalization. 
Because $\ket{h}_\theta$ is the eigenstate of $\mathcal{L}_0(\theta)=\cosh(2\theta)L_0-\sinh(2\theta)\frac{L_1+L_{-1}}{2}$ with the $\theta$-independent eigenvalue $h$, this state corresponds to the zero-energy state of $\mathcal{H}_\mathrm{SSD}$ as $\theta\to\infty$. The norm of $\ket{h}_\theta$ is kept to 1 for all $\theta$. Hence one can obtain the expectation value of any observable in this state 
by computing it with $\ket{h}_\theta$ and taking the limit $\theta\to\infty$. This limiting process provides a way to compute the expectation values in the eigenstates of the SSD Hamiltonian even though almost all of them are no longer normalizable. 

\subsection{Another type of zero-energy states of SSD Hamiltonian}
From the construction, we have
\begin{equation}
\mathcal{H}_{\mathrm{SSD}}(\theta)\mathcal{L}_n(\theta)\ket{0}_\theta=\frac{\mathcal{L}_0(\theta)}{\cosh(2\theta)}\mathcal{L}_n(\theta)\ket{0}_\theta=\frac{-n}{\cosh(2\theta)}\mathcal{L}_n(\theta)\ket{0}_\theta,\ \ n\leq-2.
\end{equation}
Because $\mathcal{H}_\mathrm{SSD}(\theta\to\infty)=\mathcal{H}_\mathrm{SSD}=L_0-(L_1+L_{-1})/2$ obeys, the zero-energy states in the CFT with the SSD can be obtained as
\begin{equation}
\label{eq3.2.1}
\lim_{\theta\to\infty}\mathcal{L}_n(\theta)\ket{0}_\theta,
\end{equation}
where $n\leq-2$ is a finite integer\footnote{As seen later, the states $\mathcal{L}_n(\theta),\ n\le -2$, go to the same state as $\theta \to \infty$.}. Norms of these states are finite because ${}_\theta\braket{0|\mathcal{L}^\dag_n(\theta)\mathcal{L}_n(\theta)|0}_\theta=-(c/12)n(n^2-1)$ obeys for arbitrary $\theta$.\par
We can obtain zero-energy states of the SSD Hamiltonian, which correspond to the states (\ref{eq3.2.1}). Using Eq. (\ref{eq5}), we have
\begin{align}
\mathcal{L}_{-m}(\theta)\ket{0}_\theta &=\mathrm{e}^{-\theta(L_1-L_{-1})}L_{-m}\ket{0} \nonumber \\
&=\mathrm{e}^{tL_{-1}}\mathrm{e}^{-L_0\log[\cosh^2(\theta)]}\mathrm{e}^{-tL_1}L_{-m}\ket{0} \nonumber \\
&=\sum_{n=0}^{m-2}\sum_{k=n-(m-2)}^{\infty} (-1)^n \begin{pmatrix} m+1 \\ n \end{pmatrix} \cdot \begin{pmatrix} m-n+k-2 \\ k \end{pmatrix} \nonumber \\
&\ \ \ \ \ \ \ \ \ \ \ \  \ \ \ \ \ \ \ \ \ \ \ \ \ \ \ \ \ \ \ \  \times t^{n+k} (1-t^2)^{m-n}L_{-(m-n+k)}\ket{0},
\end{align}
for $m\geq2$. Here, we have expressed each exponential term as a Taylor expansion and used formulas in Appendix. C. Putting $l=m-n+k$, we have
\begin{align}
\mathcal{L}_{-m}(\theta)\ket{0}=\sum_{l=2}^\infty \sum_{k=\max(l-m,0)}^{l-2} &\begin{pmatrix} m+1 \\ m-(l-k) \end{pmatrix} \cdot \begin{pmatrix} l-2 \\ k \end{pmatrix} \nonumber \\
&\ \ \ \ \ \ \ \ \ \times (-1)^{m-(l-k)}t^{m+l-2(l-k)}(1-t^2)^{l-k}L_{-l}\ket{0}.
\end{align}
The states (\ref{eq3.2.1}) are obtained by taking the limit $\theta\to\infty$. For example, we have
\begin{equation}
\label{L-2}
\lim_{\theta\to\infty}\mathcal{L}_{-2}(\theta)\ket{0}_\theta=\lim_{\theta\to\infty} \frac{4}{\sinh^2(2\theta)}\sum_{l=2}^\infty t^{l}L_{-l}\ket{0}.
\end{equation}
This state is consistent with the zero-energy state $\sum_{n>1} L_{-n}\ket{0}$ found in Ref. \cite{Tada}, and the state $\sum_{l} t^{l}L_{-l}\ket{0}$ turns out to diverge in the order of $\sinh^2(2\theta)$ as $\theta\to\infty$. Furthermore, we obtain 
\begin{align}
\label{L-3}
\lim_{\theta\to\infty}\mathcal{L}_{-3}(\theta)\ket{0}_\theta&=-\lim_{\theta\to\infty}\frac{16t}{\sinh^2(2\theta)}\sum_{l=2}^\infty t^{l}L_{-l}\ket{0}+\frac{8}{\sinh^3(2\theta)}\sum_{l=2}^\infty (l-2)t^l L_{-l}\ket{0},
\end{align}
but this state has the same corresponding state in the CFT with the SSD as ${\displaystyle \lim_{\theta\to\infty} \mathcal{L}_{-2}(\theta)\ket{0}}$ because the contribution of the second term in the RHS of Eq. (\ref{L-3}) goes to zero as $\theta\to\infty$. \par
More generally, the Virasoro operators become dependent on each other in the limit $\theta\to\infty$. This fact can be seen from the relation
\begin{equation}
\label{dependent}
\lim_{\theta\to\infty} [\mathcal{L}_n(\theta)-(-1)^{n-m}\mathcal{L}_m(\theta)]=-\frac{n-m}{2}(L_1-L_{-1}),
\end{equation}
which is proved in Appendix D. 
It is clear from the above relation that the states $\mathcal{L}_n(\theta)\ket{0},\ n\le -2$, converge to the same state since the vacuum state $\ket{0}_\theta=\ket{0}$ annihilates $L_{1}-L_{-1}$. Therefore, the states ${\displaystyle \lim_{\theta\to\infty} \mathcal{L}_n(\theta)\ket{0}}$ are the same state as long as the index $n$ is kept to a finite integer. \par
When we expand $\mathcal{L}_n(\theta)$ by conventional Virasoro generators, the coefficient of $L_m$ converges to zero as $\theta\to\infty$ for $m\le -2$ and the expansion is not valid for $\theta\to\infty$. In spite of this fact, the state $\mathcal{L}_n(\theta)\ket{0}$ has a finite norm for all $\theta$. Therefore, this regularized state gives a way to compute expectation values in the corresponding zero-energy state $\sum_l L_l\ket{0}$, as well as the primary state (\ref{primary}).

\section{Application to other deformations}
In the previous section, we obtained the zero-energy states of the CFT with SSD by considering the M$\mathrm{\ddot{o}}$bius quantization as a unitary transformation on the Virasoro algebra. We can also obtain zero-energy states in the CFT with the ``{\it k}-th angle SSD''.\par
Let us consider the Hamiltonian 
\begin{equation}
\mathcal{H}_{k{\rm -}\mathrm{SSD}}=L_0-\frac{L_k+L_{-k}}{2}+\bar{L}_0-\frac{\bar{L}_k+\bar{L}_{-k}}{2}.
\end{equation}
Note that we consider only the holomorphic part in the following. 
This CFT Hamiltonian corresponds to the Hamiltonian of the 1D lattice system whose size is $N$, i.e.
\begin{equation}
H_k=\sum_{x=1}^{N} \sin^2\left( \frac{k\pi x}{N}\right) h_{x,x+1}+\sum_{x=1}^{N} \sin^2\biggl[ \frac{k\pi}{N}\left( x-\frac{1}{2} \right)\biggr] h_x.
\end{equation}
Here $h_{x,x+1}$ and $h_x$ are local Hamiltonians. Putting $\lambda_k=\mp t/k$, $\lambda_{-k}=\pm t/k$, and $\lambda_0=-\log[\cosh^2(\theta)]/k$ where $t:=\tanh (\theta)$ in Eq. (\ref{eq4}), we have
\begin{equation}
\label{eq4.2}
\mathrm{e}^{tL_{-k}/k}\mathrm{e}^{-L_0\log[\cosh^2(\theta)]/k}\mathrm{e}^{-tL_k/k}=\exp\biggl\{ -\frac{\theta}{k}(L_k-L_{-k})+\frac{c}{24}\frac{k^2-1}{k}\log[\cosh^2(\theta)]\biggr\}.
\end{equation}
Now, we introduce transformed Virasoro generators as 
\begin{equation}
\label{eq4.1}
\mathcal{L}^{(k)}_n(\theta)=\mathrm{e}^{-\theta (L_k-L_{-k})/k}L_n\mathrm{e}^{\theta (L_k-L_{-k})/k}.
\end{equation}
We can derive a differential equation for $\mathcal{L}^{(k)}_n(\theta)$ with $n=0,\pm k$ by differentiating Eq. (\ref{eq4.1}):
\begin{equation}
\frac{\mathrm{d}}{\mathrm{d}\theta}\begin{pmatrix} \mathcal{L}^{(k)}_k(\theta) \\ \mathcal{L}^{(k)}_0(\theta) \\ \mathcal{L}^{(k)}_{-k}(\theta) \end{pmatrix} 
=A\begin{pmatrix}
\mathcal{L}^{(k)}_k(\theta) \\ \mathcal{L}^{(k)}_0(\theta) \\ \mathcal{L}^{(k)}_{-k}(\theta) \end{pmatrix}+\bm{b},
\end{equation}
where
\begin{equation}
A:=
\begin{pmatrix}
0 & -2 & 0 \\
-1 & 0 & -1 \\
0 & -2 & 0
\end{pmatrix},\ \ \bm{b}:=
\begin{pmatrix}
-\frac{c}{12}k(k^2-1) \\ 0 \\ \frac{c}{12}k(k^2-1).
\end{pmatrix}.
\end{equation}
Using $\mathcal{L}_n^{(k)}(\theta=0)=L_n$ and
\begin{equation}
\mathrm{e}^{\theta A}=\begin{pmatrix}
\cosh^2(\theta)  & -\sinh(2\theta) & \sinh^2(\theta) \\
-\sinh(2\theta)/2 & \cosh(2\theta) & -\sinh(2\theta)/2 \\
\sinh^2(\theta) & -\sinh(2\theta) & \cosh^2(\theta) \end{pmatrix},
\end{equation}
we have
\begin{equation}
\mathcal{L}_0^{(k)}(\theta)=\cosh(2\theta)L_0-\frac{\sinh(2\theta)}{2}(L_k+L_{-k}).
\end{equation}
Therefore the $k$-th angle Hamiltonian corresponds to the limit ${\displaystyle \lim_{\theta\to\infty}\mathcal{L}_0^{(k)}(\theta)/\cosh(2\theta)}$ and we can obtain zero-energy states of $\mathcal{H}_{k{\rm -}\mathrm{SSD}}$ in the same way as those of the SSD Hamiltonian. For example, the vacuum state of $\mathcal{H}_{k{\rm -}\mathrm{SSD}}$ is defined as
\begin{align}
\lim_{\theta\to\infty}\ket{0}^{(k)}_\theta&=\lim_{\theta\to\infty}\mathrm{e}^{-\theta(L_k-L_{-k})}\ket{0} \nonumber \\
&=\lim_{\theta\to\infty}\frac{1}{[\cosh(\theta)]^{\frac{c}{12}\frac{k^2-1}{k}}}\mathrm{e}^{tL_{-k}/k}\ket{0}.
\end{align}
Here, we used Eq. (\ref{eq4.2}). This state also has a finite norm for all $\theta$, so that we can calculate an expectation value of a physical quantity.

\section{Conclusion}
We have discussed the properties of two-dimensional unitary CFTs with the SSD. We showed that the Verma module can be decomposed into $SL(2,\mathbb{R})$ invariant subspaces and derived a representation of the SSD Hamiltonian in each subspace. 
Using this representation, we showed that the SSD Hamiltonian have no normalizable eigenstate except the vacuum state $\ket{0}$ annihilated by both $L_{\pm 1}$. 
We then showed that the M$\mathrm{\ddot{o}}$bius quantization introduced in \cite{Okunishi} can be thought of as a unitary transformation that relates the original Virasoro generators with the deformed ones via $\mathcal{L}_n(\theta)=\mathrm{e}^{-\theta(L_1-L_{-1})}L_n\mathrm{e}^{\theta(L_1-L_{-1})}$. This allows us to obtain zero-energy states of the SSD Hamiltonian in a systematic way. For instance, $\mathrm{e}^{-\theta(L_1-L_{-1})}\ket{h}$ naturally obtained from the primary state $\ket{h}$ of the original Hamiltonian becomes a zero-energy state of the SSD Hamiltonian after taking the limit $\theta \to \infty$. The regularization procedure using the M$\mathrm{\ddot{o}}$bius quantization also permits the calculation of expectation values in the zero-energy states even though they are not normalizable in the standard sense. 
Inspired by the correspondence between the M$\mathrm{\ddot{o}}$bius quantization and the unitary transformation by $L_1 - L_{-1}$, we have introduced a generalization of the SSD, which we call the $k$-th angle SSD. We showed that the zero-energy state of the $k$-th SSD Hamiltonian can again be obtained by using the corresponding unitary transformation. Though we have focused on the properties of the zero-energy states throughout the paper, it would be interesting to see in future studies if a regularization and limiting procedure similar to the one developed in this paper applies to non-zero energy states.

\section*{Acknowledgements}
The authors thank Kouichi Okunishi and Tsukasa Tada for helpful discussions. H. K. was supported in part by JSPS KAKENHI Grant No. JP15K17719 and No. JP16H00985.

\appendix
\section{Derivation of Eq. (\ref{wuni})}
Here we show that Eq. (\ref{wuni}) is equivalent to Eq. (\ref{wmobius}). From the definition of the Witt algebra, $\ell_n = -z^{n+1} \partial_z$, we have
\begin{align}
\mathrm{e}^{-t\ell_1}\mathrm{e}^{\ell_0\log[\cosh^2(\theta)]}\mathrm{e}^{t\ell_{-1}}z
&=\mathrm{e}^{-t\ell_1}\mathrm{e}^{\ell_0\log[\cosh^2(\theta)]}\sum_{n=0}^\infty \frac{(-t)^n}{n!}\frac{\mathrm{d}^n}{\mathrm{d}z^n}z \nonumber \\
&=\mathrm{e}^{-t\ell_1}\mathrm{e}^{\ell_0\log[\cosh^2(\theta)]}z-t \nonumber \\
&=\mathrm{e}^{-t\ell_1}\sum_{n=0}^\infty \frac{\{-\log[\cosh^2(\theta)] \}^n}{n!}\left( z\frac{\mathrm{d}}{\mathrm{d}z}\right)^nz-t \nonumber \\
&=\frac{1}{\cosh^2(\theta)}\mathrm{e}^{-t\ell_1}z-t \nonumber \\
&=\frac{1}{\cosh^2(\theta)}\sum_{n=0}^\infty t^nz^{n+1}-t \nonumber \\
&=\frac{1}{\cosh^2(\theta)}\frac{z}{1-tz}-t.
\end{align}

\section{Derivation of Eq. (\ref{eq8})}
It was shown in Ref. \cite{Okunishi} that the M$\mathrm{\ddot{o}}$bius quantization acts on the Virasoro operators as 
\begin{align}
\label{A1}
\mathcal{L}_n(\theta)&=(-1)^{n+1}\frac{\sinh(2\theta)}{2t^n}\sum_m L_m \oint_\tau \frac{\mathrm{d}z}{2\pi\mathrm{i}}\frac{(z-t)^{n+1}}{(z-1/t)^{n-1}}z^{-m-2} \nonumber \\
&=(-1)^{n+1}\frac{\sinh(2\theta)}{2}\sum_m C_m^n(t)L_m,
\end{align}
and the coefficients $C_m^n(t)$ for $n>1$ are given by
\begin{equation}
\label{A2}
C_m^n(t)=\begin{cases}
(-1)^{m+1}\frac{(n+1)!}{(m+1)!(n-m)!}F(n-1,-m-1;n-m+1;t^2)t^{n-m-1}, & n\geq m\geq -1 \\
(-1)^{n+1}\frac{(m-2)!}{(n-2)!(m-n)!}F(-n-1,m-1;m-n+1;t^2)t^{m-n-1}, & n<m
\end{cases}.
\end{equation}
Moreover, $\mathcal{L}_n(\theta) (n=0,\pm 1)$ are expanded as
\begin{eqnarray}
\begin{split}
\label{A3}
\mathcal{L}_{0}(\theta)&=-\frac{\sinh(2\theta)}{2}\biggl[L_1+L_{-1}-\left( t+\frac{1}{t} \right)L_0 \biggr], \\
\mathcal{L}_{\pm 1}(\theta)&=\frac{\sinh(2\theta)}{2}\left(t^{\mp 1}L_1+t^{\pm 1}L_{-1}-2L_0 \right).
\end{split}
\end{eqnarray}

  \par
We can show that Eqs. (\ref{A1})-(\ref{A3}) are equal to Eq. (\ref{eq8}). First, suppose $n\geq m \geq -1$. Using the formula
\begin{equation}
F(a,b;c;x)=\sum_{k=0}^\infty \frac{(a)_k(b)_k}{k!(c)_k}x^k
\end{equation}
and $(-m-1)_k=0\ (k>m+1)$, we have
\begin{equation}
C_m^n(t)=(-1)^{m+1}\frac{(n+1)!}{(m+1)!(n-m)!}\sum_{k=0}^{m+1} \frac{(n-1)_k(-m-1)_k}{k!(n-m+1)_k}t^{n+2k-m-1}.
\end{equation}
Hence, taking $k=m+1-2j$ and 
\begin{align}
(-m-1)_{m+1-j}&=\frac{(-1)^{m+1}(m+1)!}{(-1)^jj!},\\
(n-m+1)_{m+1-j}&=\frac{(n+1)!}{(n-m)!}(n-j+2)_j
\end{align}
lead to
\begin{equation}
C_m^n(t)=\sum_{j=0}^{m+1} (-1)^j \frac{(n-j+2)_j(n-1)_{m-j+1}}{j!(m-j+1)!}t^{n-2j+m+1}.
\end{equation}
This is nothing but the coefficient of Eq. (\ref{eq8}) in the form of Eq. (\ref{A1}). Similarly, when $n<m$, using
\begin{align} 
(-n-1)_k&=0,\ \  k>n+1, \\
(-n-1)_{n+1-j}&=\frac{(-1)^{n+1-j}(n+1)!}{j!},\\ 
(m-2)!(m-1)_{n+1-j}&=(n-2)!(n-1)_{m-j+1}, \\
(n+1-j)!&=\frac{(n+1)!}{(n-j+2)_j},
\end{align}
we have
\begin{equation}
C_m^n(t)=\sum_{j=0}^{n+1} (-1)^j \frac{(n-j+2)_j(n-1)_{m-j+1}}{j!(m-j+1)!}t^{n-2j+m+1}.
\end{equation} 
By noting the definition $(0)_0:=1$, we can confirm that Eq. (\ref{eq8}) satisfies Eqs. (\ref{A3}) by direct calculation. Therefore, Eq. (\ref{eq8}) is equal to  Eqs. (\ref{A1})-(\ref{A3}).

\section{Some formulas}
Using $[AB,C]=A[B,C]+[A,C]B$, we have
\begin{align}
L_1^nL_{-m}\ket{0}&=[L_1^n,L_{-m}]\ket{0}=(m+1)L_1^{n-1}L_{-(m-1)}\ket{0} \nonumber \\
&=(m+1)mL_1^{n-2}L_{-(m-2)}\ket{0}=\cdots \nonumber \\
&=(m+2-n)_{n}L_{-(m-n)}\ket{0}, \\
L_{-1}^kL_{-(m-n)}\ket{0}&=(m-n-1)L_{-1}^{k-1}L_{-(m-n+1)}\ket{0}=\cdots \nonumber \\
&=(m-n-1)_{k}L_{-(m-n+k)}\ket{0}.
\end{align} 
We use these formulas in Sec. 4.

\section{Dependence of Virasoro generators $\mathcal{L}_n(\theta)$ with $\theta\to\infty$}
In this section, we prove the relation (\ref{dependent}). Put $S_n^{(m)}(\theta):=\frac{\sinh(2\theta)}{2}\sum_k t^ky_n^{(m,k)}$ and $T_n^{(m)}(\theta):=\sum_k kt^ky_n^{(m,k)}$. Taking the sum of Eq. (\ref{eq14})$\times t^k$ about $k$, we have
\begin{equation}
\label{eq3.2.2}
t(n-1)S_{n+1}^{(m)}(\theta)-(1+t^2)S_n^{(m)}(\theta)-t(n+1)S_{n-1}^{(m)}(\theta)=2tT_n^{(m)}(\theta).
\end{equation}
The sum $\sum_k kt^ky_n^{(m,k)}$ consists of a finite set of terms whose values are nonvanishing because Eq. (\ref{eq8}) ensures that $y_n^{(m,k)}$ is zero for large $k$. Therefore, this sum converges and we have ${\displaystyle \lim_{\theta\to\infty}T_n^{(m)}(\theta)=\sum_{k} \lim_{\theta\to\infty}kt^ky_n^{(m,k)}=\sum_k ky_n^{(m,k)}=:{T'}_n^{(m)}}$. Taking the sum of Eq. (\ref{eq14})$\times (k-1)$, we obtain
\begin{align}
\label{T'}
(n-1)&{T'}_{n+1}^{(m)}-(n+1){T'}_n^{(m)}\nonumber \\
&=\sum_{k=-\infty}^{\infty} \{(k+1)(k-1)y_n^{(m,k)}-[(k-2)+1][(k-2)-1]y_n^{(m,k-2)}\}=0.
\end{align}
It follows from Eq. (\ref{eq8}) and direct calculation that  ${T'}_1^{(m)}={T'}_2^{(m)}=0$. The relation ${T'}_{-n}^{(-m)}={T'}_n^{(m)}$ follows from $y_{-n}^{(-m,k)}=y_n^{(m,k)}$, and hence we have ${T'}_n^{(m)}=0$ for all $n$ and $m$ from Eq. (\ref{T'}). Using ${\displaystyle T_n^{(m)}(\theta\to\infty)={T'}_n^{(m)}=0}$, Eq. (\ref{eq3.2.2}) leads to
\begin{equation}
\label{eq3.2.3}
(n-1)S_{n+1}^{(m)}(\theta\to\infty)-2S_n^{(m)}(\theta\to\infty)-(n+1)S_{n-1}^{(m)}(\theta\to\infty)=0.
\end{equation}
From Eqs. (\ref{e.g.}) and (\ref{A3}), we have $S_n^{(m)}(\theta\to\infty)=0$ for $m\geq 2$ and $n=-1,0,1,2$, so that Eq. (\ref{eq3.2.3}) gives $S_n^{(m)}(\theta\to\infty)=0$ for $m\geq 2$ and all $n$.\par
On the other hand, the coefficients of $L_{\pm 1}$ and $L_0$ diverge as $\theta\to\infty$. However, using Eq. (\ref{eq8}), we obtain
\begin{eqnarray}
\begin{split}
\mathcal{L}_n(\theta)=(-1)^{n+1}\frac{\sinh(2\theta)}{2}\biggl\{ t^n&L_{-1}+[(n-1)t^{n+1}-(n+1)t^{n-1}]L_0\\
&+\biggl[ \frac{n(n-1)}{2}t^{n+2}-(n^2-1)t^n+\frac{n(n+1)}{2}t^{n-2} \biggr]L_1 \biggr\} +\cdots,
\end{split}
\end{eqnarray}
which leads to
\begin{equation}
\lim_{\theta\to\infty} [\mathcal{L}_n(\theta)-(-1)^{n-m}\mathcal{L}_m(\theta)]=-\frac{n-m}{2}(L_1-L_{-1}).
\end{equation}

\end{document}